\documentclass{aa}  

\usepackage{natbib}
\usepackage{graphicx}
\usepackage{txfonts}
\usepackage{hyperref}

\begin{document} 

        \title{New ex-OH maser detections in the northern celestial
        hemisphere}
\author{O. Patoka\inst{2}\fnmsep\inst{1}, O. Antyufeyev\inst{2}\fnmsep\inst{1}, I. Shmeld\inst{1}, V. Bezrukovs\inst{1}, M. Bleiders\inst{1}, A. Orbidans\inst{1}, A. Aberfelds\inst{1} \and V. Shulga\inst{3}\fnmsep\inst{2}}

        \institute{Engineering Research Institute "Ventspils International Radio Astronomy Centre", Ventspils University of Applied Sciences, Inženieru Iela. 101, Ventspils, LV-3601, Latvia\\
                \email{artis.aberfelds@venta.lv}
                \and
                Institute of Radio Astronomy of the National Academy of Sciences of Ukraine 
                \and
                College of Physics and International Center of Future Science of Jilin University, Qianjin Street 2699, Changchun, 130012, People’s Republic of China
        }

        \abstract
        {}
        {Molecular masers, including methanol and hydroxyl masers, and in particular the ones in excited rotational states (ex-OH masers), are one of the most informative tools for studying star-forming regions. So, the discovery, of new maser sources in these regions is of great importance. Many studies and surveys of ex-OH maser sources have been carried out in the southern celestial hemisphere, but only a few have been done in the northern hemisphere. The specific aim of this work is to close this gap.}
        {The star-forming regions in the northern hemisphere with known active methanol masers were observed to search for new ex-OH maser sources with the 32 m and 16 m radio telescopes of the Ventspils International Radio Astronomy Centre (VIRAC).}
        {Three OH maser lines in the excited state at the 6035 MHz in three northern hemisphere star-forming regions are detected. The maser 189.030+0.783 was previously known, but we suggest this maser is a possible variable. We confirm recent detections of the ex-OH masers 85.41+0.00 and 90.92+1.49 by other authors. The magnetic field strength in the masering regions is estimated by using right circular polarization (RCP) and left circular polarization (LCP) pair splitting. The high-velocity resolution provides us with an estimation of a comparatively small magnetic field strength for the 189.030+0.783 and 90.92+1.49 star-forming regions.}
        {}
        
\authorrunning{O. Patoka et al.}
        \maketitle
        \section{Introduction}
        Hydroxyl and methanol molecules, and particularly their maser emission, are among the most informative for studying star-forming regions by the number of rotational-vibrational transitions available for observation in the microwave and submillimeter ranges. The most spectacular of them in the centimeter range and widely used for star formation region research are OH ground-state masers in the 1.6 GHz range first observed in the 60s (Gray \citeyear{gray_2012}) and a methanol maser line at 6.67 GHz discovered by Menten (\citeyear{1991ApJ...380L..75M}). The first OH maser in the excited rotational state (ex-OH) at the emission frequency of 6035 MHz was detected in 1969 (Yen \citeyear{Yen}). In addition to the ground levels of OH radiation, they also play an important and unique role in  the study of the star-forming regions. One of the reasons is that not all ex-OH masers have a 1665 MHz line counterpart (Avison \citeyear{Avison_2016}), and even if they are both present the higher rotational-state masers in OH correspond to physical conditions that are denser and warmer (Gray \citeyear{gray_2012}). Therefore, information from several maser lines must be used, if available, to map a region that varies considerably in density (Gray \citeyear{gray_2012}). Generally speaking, the observational differences among the various OH maser lines may help constrain maser models and thereby provide additional insight as to the details of the physical processes involved in OH maser activity (Fish and Sjouwermann \citeyear{Fish_and_Sjouwermann_2007}). 

The circular polarization of OH masers, and in particular ex-OH masers, makes it possible to recognize the probable Zeeman pairs. This makes them a good tool for studying magnetic fields in the regions where they are located (Gray \citeyear{gray_2012}). Ground-level OH maser lines have a higher Lande factor and thus a larger Zeeman split, and one may consider that their use for magnetic field measurements would be preferable to ex-OH maser radiation. However, for the reasons mentioned above, this radiation is also important for these measurements, and in many cases it would even be unique in its ability to measure magnetic fields from which OH 1.6 GHz maser radiation does not emanate. In addition, there are indications that in some cases the Zeeman components of this radiation are more difficult to identify as ex-OH maser radiation (Cook \citeyear{Cook_(1977)}, Crutcher and Kemball \citeyear{Crutcher_and_Kemball_2019}).

        Baudry \citeyear{Baudry} observed 168 sources of the northern celestial hemisphere at frequencies of 6030 and 6035 MHz toward intense IRAS point sources. These authors detected ex-OH masers emission in 16\% of the total number of observed sources, and investigated magnetic fields in the Galaxy plane. Fish \citeyear{Fish} carried out the survey of the already known ex-OH masers. The studies of single objects have also been carried out repeatedly in the northern celestial hemisphere (e.g., Al-Marzouk \citeyear{Al-Marzouk}, Asanok \citeyear{Asanok}, Etoka \citeyear{Etoka}, Fish and Sjouwerman \citeyear{Fish-Sjouwerman}). Many studies and surveys have been carried out in the southern celestial hemisphere (e.g., Caswell \citeyear{Caswell2003}, Caswell \citeyear{Caswell2001}, Caswell \citeyear{Caswell1997}, Knowles \citeyear{Knowles}, Avison \citeyear{Avison_2016}). However, the opposite is true with regard to ex-OH surveys in the northern sky. In this work, we aim to close this gap. 
        
        In 2018, we started an observational campaign with goal to search for radiation from ex-OH masers in the northern celestial hemisphere with a high-frequency resolution to detect new excited-state OH sources, to compare with previously known sources to find the possible variability, and to find the most interesting sources for future interferometric observations.
        
        In September 2020,  while this paper was under review, Szymczak et al. \citeyear{torun_exoh} published results similar to those of our project. These authors carried out observations toward the 445 methanol masers from the Torun catalog (Szymczak et al. \citeyear{Szymczak}) and the Multibeam Methanol Survey (Green et al. \citeyear{Green_2010}, Breen et al. \citeyear{Breen_2015}). Our source sample is smaller and contains most of the sources from the Szymczak survey and previously unobserved sources at 6035 MHz from the Yang et al. \citeyear{Yang2019} sample. However, our observations were carried out with a five-times-better velocity resolution
        \citep[0.1 km s$^{-1}$ for][and 0.019 km s$^{-1}$ in this paper]{torun_exoh}
        and slightly better sensitivity.
        
        In this paper, we describe our source selection criteria, observation technique, and three  detected masers (two of them were unknown to us when we first submitted this paper)   at 6035 MHz.
        After the first submission of this paper, we also noted seven possible newly detected masers at 6035 MHz that were not observed, or observed but not detected, by Szymczak et al. \citeyear{torun_exoh} or other authors. However to confirm these new detections, some additional observations and literature studies for possible previous detections  are necessary (see Section 4, below). We intend to present the results of these efforts and the full survey in the subsequent paper.

        \section{Source selection}

        Simulations of ex-OH masers show that emission at the 6031 and 6035 MHz transitions should originate in regions with low gas temperatures (T$_K$ <70 K) and high densities (up to $n_{\rm H}$ = 10$^{8.5}$ ${cm^{-3}}$) (Cragg \citeyear{Cragg}). The physical conditions under which ex-OH masers originate overlap with the 6.7 GHz methanol masers, but the detailed parameters are not identical (Gray et al. \citeyear{Gray_1992}). This means that at least some of the ex-OH masers should exist in the regions with class II methanol masers. The ex-OH maser pumping mechanism is mainly radiative. Therefore, these masers should be associated with the infrared sources and with the ultra-compact (UC) HII regions. (Fish \citeyear{Fish_2007IAUS} and references therein). At the same time, methanol masers at 6.7 GHz are associated with regions of massive star formation (e.g., Gómez-Ruiz et al. \citeyear{Gomez-Ruiz_2016}), infrared sources (e.g., Sun et al. \citeyear{Sun_2014}), young stellar objects (e.g., Gómez-Ruiz et al. \citeyear{Gomez-Ruiz_2016}), and UC HII regions (e.g., Hu et al. \citeyear{Hu_2016}). Therefore, we took methanol maser positions as target points for our observations.

        We prepared the input source list for our observational program, which contains a total of 272 sources. The coordinates and $V_{lsr}$ we selected from the positions of the methanol masers 6.7 GHz (Yang et al. \citeyear{Yang2019}, Fontani et al. \citeyear{Fontani_2010}, Szymczak et al. \citeyear{Szymczak}) and most precise known positions of methanol masers were used. The masers in these catalogs were detected toward the WISE point sources, YSOs, and IRAS point sources. From these catalogs, we only took sources with possible elevation higher than about 25 degrees with regard to the Ventspils telescope site (this corresponds to the declination -7.5 degrees). Our telescope operation at lower degrees is not so good due to several technical limitations. From these catalogs, we only selected sources which to have separation more than 3' on the sky based on the full width at half maximum (FWHM) of the main beam for a 32 meter radio telescope (~6').
        
        Some of the sources from our source list have previously been observed (e.g., Baudry et al. \citeyear{Baudry}, Avison et al. \citeyear{Avison_2016}, Szymczak et al. \citeyear{torun_exoh}). These previously observed sources also allow the comparative analysis and verification of long-term variations in masers. The full source list will be published in a subsequent paper after observations have ended.
        
        
        \section{Observations}
        The observation campaign started in March 2018. For observations, we mainly use the Irbene RT-32 radio telescope of the Ventspils International Radio Astronomy Centre. For equipment tests or detection confirmations, we also used the RT-16 radio telescope. Sources described in this paper were observed with the RT-32.
        
        We used cryogenic broadband receivers on both telescopes for the frequency range of 4.5 – 8.8 GHz developed and installed by the Tecnologias de Telecomunicaciones e Informacion (TTI\footnote{http://www.ttinorte.es/}) company. Receivers incorporate noise-diode and phase-calibration tone injection for amplitude and phase calibration during observation sessions. Both receivers have identical schematics, only wideband feed horns are different and adapted for the geometry of each antenna. Both receivers simultaneously register two polarizations (left circular polarization (LCP) \& right circular polarization (RCP)) (Bezrukovs \citeyear{2020arXiv200805794B}). The calibration methods for single-dish observation were used, as described by Winkel \citeyear{Winkel_2012}. Both telescope systems were regularly calibrated, including polarization, by observing strong stable flux sources from Perley et al. \citeyear{2013ApJS..204...19P}. The system noise temperature was about 25 K for the RT-32 and  about 33 K for the RT-16. A new FFT spectrometer based on the Ettus Research USRP X300\footnote{https://kb.ettus.com/X300/X310} software-defined radio (SDR) platform was used for spectral analysis (Bleiders et al. \citeyear{Bleiders}). The spectrometer can be configured for a 12 MHz bandwidth with 16384 channels for simultaneous observations on the LCP \& RCP. Given the spectrometer bandwidth, we were able to observe at a center frequency of 6032.92 MHz, which made it possible to observe radiation at two transitions of the OH molecule simultaneously: $\prod_{3/2}^{}$ $J=5/2,$ $F=2-2$ (6030.747 MHz) and  $\prod_{3/2}^{}$ $J=5/2,$ $F=3-3$ (6035.092 MHz). In this paper, we observed two excited transitions of the OH molecule at the transition frequencies 6030.747 and 6035.092 MHz. With an rms above 0.45 Jy per polarization, only 6035 MHz maser lines were detected. In these cases lacking a better frequency resolution, we were obliged to switch to a bandwidth narrower than 1.5625 MHz and a spectrometer with 4096 channels, and the total on-source time was also increased from about 2 h to 4 h in order to obtain the better signal-to-noise ratio. All observations were carried out in the frequency switching mode with a frequency throw of 0.390625 MHz. The typical rms noise level of our observations after four hours of integration time was 0.1-0.2 Jy for a single polarization flux density.
        
        Our developed Python scripts were used for folding frequency switching and the averaging of scans. The GILDAS package\footnote{http://iram.fr/IRAMFR/GILDAS/} was used to subtract the baseline and line-parameter estimation. Channel separation was equal to 0.381 kHz, corresponding to 0.019 km s$^{-1}$.

        
        \section{Results}
        
        At the time this paper was submitted, we had observed 78 sources from our input catalog at 6035 MHz. Based on the S/N > 3 criteria, we detected 16 ex-OH masers at a frequency of 6035 MHz. Three of them are described in this paper. However, the remaining 13 are to be confirmed and processed more accurately. They will be described in a forthcoming paper. 
        
        In this paper, we describe three detected ex-OH masers at 6035 MHz from our detected sample of 16 sources. Two of them are new ex-OH masers at 6035 MHz, but after this paper was submitted detections of these masers were reported in Szymczak et al. \citeyear{torun_exoh}. 
        
        A list of the three of our confirmed detections is given in Table~\ref{T1}. In Table 1, the source name, equatorial coordinates (J2000), observation epoch, line frequency, velocity of rest, and noise levels for each polarization are given. In Table~\ref{T3}, non-detections are given. 
        
        Standard Gaussian profiles were fit in order to retrieve the line parameters. The RCP and LCP pairs were used to estimate the magnetic field strength. The splitting factor used was 0.056 km s$^{-1}$mG$^{-1}$ for 6035 MHz (Green \citeyear{Green}, Caswell \citeyear{Caswell2003}). Results for 189.030+0.783, 85.41+0.00, and 90.92+1.49 are shown in the Table~\ref{T2}.  In this table, we give the number of the source from Table~\ref{T1}, the polarization, and Gaussian line parameters (center velocity of line, FWHM, and peak and integrated flux densities). The magnetic field strengths are also given in the Table~\ref{T2}.
        
\begin{figure}
                \centering%
                \includegraphics[width=8cm]{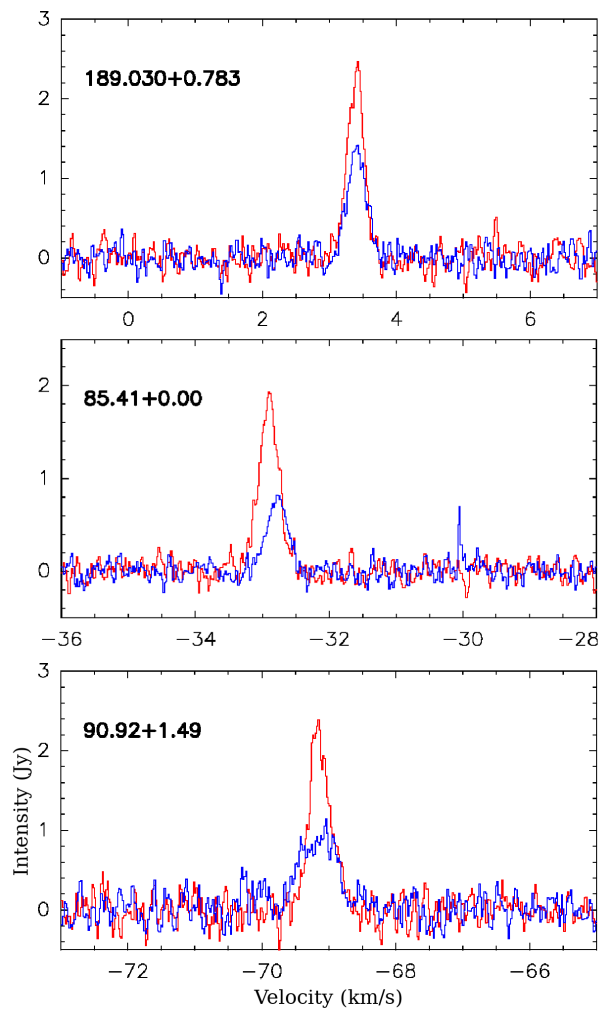}
                \caption{6035 MHz OH maser spectra detected during VIRAC observations. LCP and RCP polarizations depicted with blue and red lines, respectively.}
                \label{Fig1}
        \end{figure}

    The spectra of three our confirmed sources are shown in Fig.~\ref{Fig1}. Blue corresponds to the LCP, and red corresponds to RCP. Maser lines were not detected at 6031 MHz from these sources.

        \begin{table*}
                \caption{List of sources.}
                \label{T1}
                $$ 
                \begin{tabular}{lp{0.15\linewidth}llllp{0.09\linewidth}lp{0.1\linewidth}ll}
                \hline
                \noalign{\smallskip}
                No & Source Name &  RA (J2000)   & Dec (J2000) & Epoch   &  Trans.  & V$_{lsr}$   &  rms(LCP) & rms(RCP) \\
                   &             & (hh:mm:ss) &(dd:mm:ss)& yy/mm/dd & (MHz) & (km s$^{-1}$)&   (mJy)    & (mJy) \\
                \noalign{\smallskip}
                \hline
                \noalign{\smallskip}
                1 & 189.030+0.783   & 06:08:40.65 & 21:31:07.00 & 2020/01/16  & 6035   & 8.9    & 125  &  142 \\
                2 & 85.41+0.00  & 20:54:13.70 & 44:54:06.80 & 2020/01/17 & 6035   & -28.53 & 85  &  92 \\
                3 & 90.92+1.49  & 21:09:12.60 & 50:01:02.90 & 2019/12/24 & 6035  & -70.42 & 161   & 164 \\
                \noalign{\smallskip}
                \hline
                \end{tabular}
                $$ 
        \end{table*}

        %
        \begin{table*}
                \caption{Line parameters. Parameters without asterisks are our data, one asterisk denotes Szymczak et al. \citeyear{torun_exoh}, and two asterisks denote Avison et al. \citeyear{Avison_2016}. 1: 189.030+0.783, 2: 85.41+0.00, 3: 90.92+1.49.}
                \label{T2}
                \begin{tabular}{llp{0.2\linewidth}lllll}
                \hline
                \noalign{\smallskip}
            No  &  Pol.  &  Peak Velocity        &  FWHM  &  S$_p$ &  S$_i$  &  B \\
                &       &  (km s$^{-1}$)   &  (km s$^{-1}$) &  (Jy) &  (Jy km s$^{-1}$)  &  (mG)\\
                \noalign{\smallskip}
                \hline
                \noalign{\smallskip}
                1   & LCP & 3.413; 3.37\textsuperscript{*}; 3.36\textsuperscript{**} & 0.33  & 1.36; 1.74\textsuperscript{*}; 0.6\textsuperscript{**}  & 0.48 & -0.27\\
                & RCP & 3.398; 3.32\textsuperscript{*}; 3.27\textsuperscript{**} & 0.295 & 2.33; 2.43\textsuperscript{*}; 0.98\textsuperscript{**} & 0.73 & \\
                2  & LCP & -32.775; -32.92\textsuperscript{*} & 0.360  & 0.797; 0.74\textsuperscript{*} & 0.307 & -2.21\\
                & RCP & -32.899; -32.97\textsuperscript{*} & 0.338 & 1.823; 2.0\textsuperscript{*} & 0.667 & \\
                3  & LCP & -69.150; -69.24\textsuperscript{*} & 0.640  & 0.90; 1.0\textsuperscript{*} & 0.62 & 0.25\\
                & RCP & -69.136; -69.20\textsuperscript{*} & 0.390 & 2.17; 1.89\textsuperscript{*} & 0.88 & \\
                \noalign{\smallskip}
                \hline
                \end{tabular}
        \end{table*}
        
        \begin{table}
                \caption{List of non-detections above rms = 0.45 Jy per polarization. The central velocity for observations is also given.}
                \label{T3}
                \begin{tabular}{llp{0.2\linewidth}lll}
                
                \hline
                \noalign{\smallskip}
                 Source Name (l b) &  RA   & Dec  & V$_{lsr}$ \\
                    & (hh:mm:ss) &(dd:mm:ss)&  (km s$^{-1}$) \\
                \noalign{\smallskip}
                \hline
                \noalign{\smallskip}
                121.298+0.659 & 00:36:47.353 & 63:29:02.16 & -23.31 \\
                123.066-6.309 & 00:52:24.196 & 56:33:43.17 & -29.36 \\
                136.84+1.15 & 02:49:29.800 & 60:47:29.50 & -45.24 \\
                168.06+0.82 & 05:17:13.300 & 39:22:14.00 & -16.2 \\
                174.201-0.071 & 05:30:48.015 & 33:47:54.61 & 1.48 \\
                173.482+2.446 & 05:39:13.059 & 35:45:51.29 & -12.98 \\
                173.71+2.35 & 05:39:27.600 & 35:30:58.00 & -14.0 \\
                173.70+2.89 & 05:41:37.400 & 35:48:49.00 & -23.8 \\
                189.471-1.216 & 06:02:08.370 & 20:09:20.10 & 18.8 \\
                189.778+0.345 & 06:08:35.280 & 20:39:06.70 & 5.7 \\
                188.946+0.886 & 06:08:53.320 & 21:38:29.10 & 10.8 \\
                188.794+1.031 & 06:09:06.960 & 21:50:41.30 & -5.5 \\
                192.600-0.048 & 06:12:53.990 & 17:59:23.70 & 4.6 \\
                196.454-1.677 & 06:14:37.030 & 13:49:36.60 & 15.2 \\
                27.220+0.260  & 18:40:03.72  & -04:57:45.6 & 9.19 \\
                27.222+0.136  & 18:40:30.55 & -05:01:05.4 & 118.8 \\
                26.53-0.27    & 18:40:40.2   & -05:49:12.8 & 104.23 \\
                27.01-0.04 & 18:40:44.8 & -05:17:09.1 & -21.09 \\
                27.784+0.057 & 18:41:49.58 & -04:33:13.8 & 111.87 \\
                27.365-0.166 & 18:41:51.06 & -05:01:42.8 & 99.74 \\
                37.430+1.518 & 18:54:14.230 & 04:41:41.10 & 41.2 \\
                56.963-0.234 & 19:38:17.100 & 21:08:05.40 & 29.9 \\
                57.610+0.025 & 19:38:40.740 & 21:49:32.70 & 38.9 \\
                58.775+0.644 & 19:38:49.130 & 23:08:40.20 & 33.3 \\
                59.436+0.820 & 19:39:34.200 & 23:48:25.50 & -50.2 \\
                59.833+0.672 & 19:40:59.330 & 24:04:46.50 & 38.2 \\
                59.783+0.065 & 19:43:11.250 & 23:44:03.30 & 27.0 \\
                59.498-0.236 & 19:43:42.450 & 23:20:13.80 & 27.5 \\
                59.634-0.192 & 19:43:50.000 & 23:28:38.80 & 29.6 \\
                60.57-0.19 & 19:45:52.300 & 24:17:42.60 & 3.64 \\
                62.310+0.114 & 19:48:35.350 & 25:56:41.80 & 23.6 \\
                71.52-0.38 & 20:12:57.900 & 33:30:26.50 & 10.2 \\
                78.122+3.633 & 20:14:26.044 & 41:13:33.39 & -7.75 \\
                74.098+0.110 & 20:17:56.320 & 35:55:24.30 & -0.22 \\
                75.010+0.274 & 20:19:49.290 & 36:46:09.40 & 3.36 \\
                76.093+0.158 & 20:23:23.650 & 37:35:34.30 & 4.92 \\
                78.629+0.98 & 20:27:26.800 & 40:07:50.00 & -39.0 \\
                78.89+0.71 & 20:29:24.900 & 40:11:19.20 & -7.0 \\
                79.736+0.991 & 20:30:50.670 & 41:02:27.60 & -5.47 \\
                81.794+0.911 & 20:37:47.390 & 42:38:39.00 & 7.19 \\
                82.308+0.729 & 20:40:16.720 & 42:56:28.60 & 10.3 \\
                84.951-0.691 & 20:55:32.500 & 44:06:10.20 & -36.5 \\
                89.930+1.669 & 21:04:15.410 & 49:24:27.40 & -69.9 \\
                97.52+3.17 & 21:32:13.000 & 55:52:56.00 & -71.16 \\
                94.602-1.796 & 21:39:58.260 & 50:14:20.96 & -40.84 \\
                106.80+5.31 & 22:19:18.300 & 63:18:48.00 & -2.0 \\
                107.288+5.638 & 22:21:22.500 & 63:51:13.00 & -8.53 \\
                109.871+2.114 & 22:56:17.903 & 62:01:49.65 & -3.75 \\
                109.92+1.98 & 22:57:11.200 & 61:56:03.00 & -2.4 \\
                111.26-0.77 & 23:16:10.000 & 59:55:31.30 & -38.89 \\
                \noalign{\smallskip}
                \hline
                \end{tabular}
        \end{table}

        \section{Discussion}
        
        So far, from our 78 observed methanol maser sources, the detected or possibly detected ex-OH masers have 16 sources with S/N > 3. The detection rate of at the 6035 MHz ex – OH masers is 0.21 toward class II methanol masers at 6.7 GHz. This detection rate is slightly higher than previously obtained (Baudry et al. \citeyear{Baudry}) at the same frequency (6035 MHz) for the northern celestial hemisphere toward intense IRAS point sources and almost three times higher than the detection rate toward 6.7 GHz for methanol masers (Szymczak et al. \citeyear{torun_exoh}). One of the possible explanations may be better sensitivity achieved by the longer on-source integration time. We have not detected the 6031 MHz transition in any of our observed targets, but this can be explained by the relative weakness of this transition (Cragg et. al. \citeyear{Cragg}). All three sources we describe here have a simple 6035 MHz line profile, both in RCP and LCP, with one velocity component.
        
        \subsection{\bf 189.030+0.783 (IRAS 06056+2131)}
        We detected ex-OH maser line at the only 6035 MHz  frequency in this source. The RCP component  is  more  intense  compared  to  the  LCP  component. There is no emission at 6031 MHz. The lesser upper-intensity detection limit (three times rms) during the observations for the 6031 MHz transition frequency was 0.48 Jy and 0.46 Jy for RCP and LCP, respectively. This maser is associated with the UC H II region G189.029+0.783 with a distance of 2.0 kpc (Hu \citeyear{Hu_2016}). The 6035 MHz ex-OH maser was ~5-6 km s$^{-1}$ blueshifted from the eight methanol maser spots at the 6.7 GHz detected toward this region by interferometric observations (Hu \citeyear{Hu_2016}). Besides this, the methanol masers at the 12.2 GHz (Breen \citeyear{Breen}), 36 GHz (Val'tts \citeyear{Val'tts}), 95 GHz (Yang \citeyear{Yang_2017}), and H$_{2}$O masers (Sunada \citeyear{Sunada_2007}) were detected toward this region. The source 189.030+0.783 was already observed by (Baudry et al. \citeyear{Baudry}) at the same frequency, 6035 GHz, but there were no ex-OH maser lines detected. Later, Avison et al. \citeyear{Avison_2016} detected maser line at 6035 MHz with a frequency resolution of 1.9 kHz using the Parkes radio telescope. Recently, Szymczak et al. \citeyear{torun_exoh} detected a maser line at 6035 MHz with a velocity resolution of 0.1 km s$^{-1}$ using the Torun 32 m radio telescope.
        We detected a maser line with 1.36 and 2.33 Jy peak intensities for LCP and RCP, respectively, which is about the same as in Torun observations and approximately two times higher than in Parks observations. The peak line velocities in our observations were 3.413 and 3.398 km s$^{-1}$ for LCP and RCP. Szymczak et al. \citeyear{torun_exoh} did not estimate the magnetic fields since the relative velocity difference between LCP and RCP was too small, given their coarser velocity resolution, in order to reliably determine a B-filed strength from the $\Delta (v)$ of the polarized line components. The velocity resolution of our observations allows us to determine small velocity shifts between the polarization components. We estimated the magnetic field strength is -0.27 mG. Green et al. \citeyear{Green} carried out the interferometric ex-OH maser observations with the velocity resolution comparable with our observations. They show that the field strengths vary between 0.2 and 11.4 mG, with a median of 3.9 mG among their sample detections. Our result is close to the lower limit of the distribution, but it could be a velocity resolution limitation effect. Darwish et al. \citeyear{2020MNRAS.499.1441D} carried out high-angular-resolution observations of OH maser emission in the ground state and only detected emission from the 1665 MHz transition with peak velocities of 9.981 and 9.139 km s$^{-1}$ for LCP and RCP, respectively. Based on the LCP and RCP velocity splitting, Darwish et al. estimated the magnetic field strength -1.5 mG. This is different to the results in this paper, but our detected maser is ~6.5 km s$^{-1}$ blueshifted and therefore most likely arises from a different subcomponent of the source. Based on the comparative analyses above, we suppose that 189.030+0.783 is a variable source and could be a good candidate for monitoring observations. 
        
        \subsection{\bf 85.41+0.00}
        To our knowledge, a previously unknown ex-OH maser emission at 6035 MHz was detected for the first time by this present study toward a known methanol 6.7 GHz maser source with a distance of 5.5 kpc (Lumsden \citeyear{Lumsden_2013}) and young stellar cluster CBJC 8 (Persi \citeyear{Persi}). The RCP component is more intense compared to the LCP component. We had no detections at 6031 MHz. The lesser upper-intensity detection limit (three times rms) during the observations for the 6031 MHz transition frequency was 0.41 Jy for RCP and LCP. The 22 methanol maser spots at the 6.7 GHz are also detected in this region by interferometric observations (Hu \citeyear{Hu_2016}), and an ex-OH maser is slightly blueshifted with respect to these 6.7 GHz maser spots. The 22 GHz H$_2$O maser was detected toward this region (Urquhart et al. \citeyear{2011MNRAS.418.1689U}). This  source was also unsuccessfully searched for 6031 and 6035 ex-OH masers from 2017 March to 2017 August by the Shanghai TianMa 65 m telescope by Ouyang (\citeyear{Ouyang}). Szymczak et al. \citeyear{torun_exoh} recently detected a maser line at 6035 MHz with a velocity resolution of 0.1 km s$^{-1}$ using the Torun 32 m radio telescope. For comparison, line parameters are shown in Table~\ref{T2}. We estimated the magnetic field strength at -2.21 based on the LCP and RCP component splitting. This is a typical magnetic field strength in the star-forming regions (e.g., Baudry et al. \citeyear{Baudry}) and close to median the value 3.9 mG (Green et al. \citeyear{Green}). Our magnetic field estimate is almost twice the value of -1.2 mG, which was provided in Szymczak et al. \citeyear{torun_exoh}. This difference cannot be explained by the lower velocity resolution of Torun observations alone, and it could be a sign of the magnetic field variations. Furthermore, 85.41+0.00 was observed with VLA in a trigger observation campaign reacting to an associated methanol maser flare (reported by the Ibaraki observatory in January 2020) at multiple maser transition frequencies (1612, 1665, and 1720 MHz OH, 6035 MHz ex-OH, 6.7 and 12.18 GHz methanol and 22 GHz water).  In any case, this maser source could be useful for the further monitoring of its possible variability.
        
        \subsection{\bf 90.92+1.49}
        The ex-OH maser emission at the 6035 MHz was detected for the first time in the present work toward the known ultra-compact H II region G090.921+1.486 with a distance of 7.8 kpc (Hu \citeyear{Hu_2016}). The RCP component  is  more  intense  compared  to  the  LCP  component. We had a negative detection at 6031 MHz. The lesser upper-intensity detection limits (three times rms) during the observations for the 6031 MHz transition frequency were 0.82 Jy and 0.85 Jy for RCP and LCP, respectively. The 19 methanol maser spots at 6.7 GHz are also detected in this region by interferometric observations (Hu \citeyear{Hu_2016}). H$_{2}$O (Sunada \citeyear{Sunada_2007}) and OH masers at the 1665 MHz (Szymczak \citeyear{Szymczak_2000}) were also detected toward this region. Szymczak et al. \citeyear{torun_exoh} detected a maser line at 6035 MHz with a velocity resolution of 0.1 km s$^{-1}$ using the Torun 32 m radio telescope.
        The maser line peak intensities for LCP and RCP detected in this work are about the same as those of Torun observations. Szymczak et al. \citeyear{torun_exoh} did not estimate the magnetic fields, since for 189.030+0.783 the relative velocity difference between LCP and RCP was too small.
        We estimated the magnetic field strength to be 0.25 mG. This result is close to the lower limit of the magnetic field distribution in the star-forming regions (Green et al. \citeyear{Green}), but this limit could be a velocity-resolution limitation effect.
        
        \section{Conclusions}
        We present some initial results of a single-dish survey to detect new ex-OH masers, and we verify the state of some already known sources. The full survey will be presented in a follow-up paper.
        We observed the previously known ex-OH maser 189.030+0.783 with a better velocity resolution. We confirm the detections of 85.41+0.00 and  90.92+1.49 that were reported in Szymczak et al. \citeyear{torun_exoh} after this paper was submitted. 
        
         Our observations with a high-velocity resolution allowed us to estimate the weak magnetic fields in the 189.030+0.783 and 90.92+1.49 masers that were not estimated early by other authors. The environment around 85.41+0.00 has a magnetic field with a typical strength of -2.21 mG. 
        
        We suggest that all the described masers are good candidates for further monitoring for possible variability. For example, 85.41+0.00 experienced a recent maser outburst in the 6.7 GHz maser line, and several interferometric follow-ups (including the 6035 MHz line) are underway (see also the related subsection of the discussion section).

        Other newly detected and reobserved sources are also potentially interesting VLBI targets  to obtain a more comprehensive YSO evolution picture. Very recently, Szymczak et. al. \citeyear{torun_exoh} reported their findings from similar research, and their team also detected signal toward 85.41+0.00 and 90.92+1.49, confirming our results in this article. Nevertheless a more direct comparison between our results and those of the Szymczak team is beyond of the scope of this paper.   
        
        \begin{acknowledgements}
                This work was supported by the ERDF project “Physical and chemical processes in the interstellar medium”, No.1.1.1.1/16/A/213. Authors are thankful A. Vasyunin and M. Gawronski for useful discussion. The authors express their gratitude to the anonymous referee for the valuable criticism, remarks and suggestions.
        \end{acknowledgements}
        
        \bibliographystyle{aa} 
        \bibliography{main} 

\begin{thebibliography}{44}
\expandafter\ifx\csname natexlab\endcsname\relax\def\natexlab#1{#1}\fi

\bibitem[{{Al-Marzouk} {et~al.}(2012){Al-Marzouk}, {Araya}, {Hofner}, {Kurtz},
  {Linz}, \& {Olmi}}]{Al-Marzouk}
{Al-Marzouk}, A.~A., {Araya}, E.~D., {Hofner}, P., {et~al.} 2012, \apj, 750,
  170

\bibitem[{{Asanok} {et~al.}(2010){Asanok}, {Etoka}, {Gray}, {Thomasson},
  {Richards}, \& {Kramer}}]{Asanok}
{Asanok}, K., {Etoka}, S., {Gray}, M.~D., {et~al.} 2010, \mnras, 404, 120

\bibitem[{{Avison} {et~al.}(2016){Avison}, {Quinn}, {Fuller}, {Caswell},
  {Green}, {Breen}, {Ellingsen}, {Gray}, {Pestalozzi}, {Thompson}, \&
  {Voronkov}}]{Avison_2016}
{Avison}, A., {Quinn}, L.~J., {Fuller}, G.~A., {et~al.} 2016, \mnras, 461, 136

\bibitem[{{Baudry} {et~al.}(1997){Baudry}, {Desmurs}, {Wilson}, \&
  {Cohen}}]{Baudry}
{Baudry}, A., {Desmurs}, J.~F., {Wilson}, T.~L., \& {Cohen}, R.~J. 1997, \aap,
  325, 255

\bibitem[{{Bezrukovs} {et~al.}(2020){Bezrukovs}, {Bleiders}, {Orbidans}, \&
  {Bezrukovs}}]{2020arXiv200805794B}
{Bezrukovs}, V., {Bleiders}, M., {Orbidans}, A., \& {Bezrukovs}, D. 2020, arXiv
  e-prints, arXiv:2008.05794

\bibitem[{Bleiders {et~al.}(2020)Bleiders, Antyufeyev, Patoka, Orbidans,
  Aberfelds, Steinbergs, Bezrukovs, \& Shmeld}]{Bleiders}
Bleiders, M., Antyufeyev, O., Patoka, O., {et~al.} 2020, Journal of
  Astronomical Instrumentation, 09, 2050009

\bibitem[{{Breen} {et~al.}(2012){Breen}, {Ellingsen}, {Caswell}, {Green},
  {Voronkov}, {Fuller}, {Quinn}, \& {Avison}}]{Breen}
{Breen}, S.~L., {Ellingsen}, S.~P., {Caswell}, J.~L., {et~al.} 2012, \mnras,
  426, 2189

\bibitem[{{Breen} {et~al.}(2015){Breen}, {Fuller}, {Caswell}, {Green},
  {Avison}, {Ellingsen}, {Gray}, {Pestalozzi}, {Quinn}, {Richards}, {Thompson},
  \& {Voronkov}}]{Breen_2015}
{Breen}, S.~L., {Fuller}, G.~A., {Caswell}, J.~L., {et~al.} 2015, \mnras, 450,
  4109

\bibitem[{{Caswell}(1997)}]{Caswell1997}
{Caswell}, J.~L. 1997, \mnras, 289, 203

\bibitem[{{Caswell}(2001)}]{Caswell2001}
{Caswell}, J.~L. 2001, \mnras, 326, 805

\bibitem[{{Caswell}(2003)}]{Caswell2003}
{Caswell}, J.~L. 2003, \mnras, 341, 551

\bibitem[{{Cook}(1977)}]{Cook_(1977)}
{Cook}, A.~H. 1977, {Celestial masers}

\bibitem[{Cragg {et~al.}(2002)Cragg, Sobolev, \& Godfrey}]{Cragg}
Cragg, D.~M., Sobolev, A.~M., \& Godfrey, P.~D. 2002, Monthly Notices of the
  Royal Astronomical Society, 331, 521

\bibitem[{{Crutcher} \& {Kemball}(2019)}]{Crutcher_and_Kemball_2019}
{Crutcher}, R.~M. \& {Kemball}, A.~J. 2019, Frontiers in Astronomy and Space
  Sciences, 6, 66

\bibitem[{{Darwish} {et~al.}(2020){Darwish}, {Richards}, {Etoka}, {Edris},
  {Saad}, {Beheary}, \& {Fuller}}]{2020MNRAS.499.1441D}
{Darwish}, M.~S., {Richards}, A.~M.~S., {Etoka}, S., {et~al.} 2020, \mnras,
  499, 1441

\bibitem[{{Etoka} {et~al.}(2012){Etoka}, {Gray}, \& {Fuller}}]{Etoka}
{Etoka}, S., {Gray}, M.~D., \& {Fuller}, G.~A. 2012, in IAU Symposium, Vol.
  287, Cosmic Masers - from OH to H0, ed. R.~S. {Booth}, W.~H.~T. {Vlemmings},
  \& E.~M.~L. {Humphreys}, 171--175

\bibitem[{{Fish}(2007)}]{Fish_2007IAUS}
{Fish}, V.~L. 2007, in IAU Symposium, Vol. 242, Astrophysical Masers and their
  Environments, ed. J.~M. {Chapman} \& W.~A. {Baan}, 71--80

\bibitem[{{Fish} {et~al.}(2006){Fish}, {Reid}, {Menten}, \& {Pillai}}]{Fish}
{Fish}, V.~L., {Reid}, M.~J., {Menten}, K.~M., \& {Pillai}, T. 2006, \aap, 458,
  485

\bibitem[{{Fish} \& {Sjouwerman}(2007)}]{Fish_and_Sjouwermann_2007}
{Fish}, V.~L. \& {Sjouwerman}, L.~O. 2007, \apj, 668, 331

\bibitem[{{Fish} \& {Sjouwerman}(2010)}]{Fish-Sjouwerman}
{Fish}, V.~L. \& {Sjouwerman}, L.~O. 2010, \apj, 716, 106

\bibitem[{{Fontani} {et~al.}(2010){Fontani}, {Cesaroni}, \&
  {Furuya}}]{Fontani_2010}
{Fontani}, F., {Cesaroni}, R., \& {Furuya}, R.~S. 2010, \aap, 517, A56

\bibitem[{{G{\'o}mez-Ruiz} {et~al.}(2016){G{\'o}mez-Ruiz}, {Kurtz}, {Araya},
  {Hofner}, \& {Loinard}}]{Gomez-Ruiz_2016}
{G{\'o}mez-Ruiz}, A.~I., {Kurtz}, S.~E., {Araya}, E.~D., {Hofner}, P., \&
  {Loinard}, L. 2016, \apjs, 222, 18

\bibitem[{Gray(2012)}]{gray_2012}
Gray, M. 2012, Maser Sources in Astrophysics, Cambridge Astrophysics (Cambridge
  University Press)

\bibitem[{{Gray} {et~al.}(1992){Gray}, {Field}, \& {Doel}}]{Gray_1992}
{Gray}, M.~D., {Field}, D., \& {Doel}, R.~C. 1992, \aap, 262, 555

\bibitem[{{Green} {et~al.}(2010){Green}, {Caswell}, {Fuller}, {Avison},
  {Breen}, {Ellingsen}, {Gray}, {Pestalozzi}, {Quinn}, {Thompson}, \&
  {Voronkov}}]{Green_2010}
{Green}, J.~A., {Caswell}, J.~L., {Fuller}, G.~A., {et~al.} 2010, \mnras, 409,
  913

\bibitem[{Green {et~al.}(2015)Green, Caswell, \& McClure-Griffiths}]{Green}
Green, J.~A., Caswell, J.~L., \& McClure-Griffiths, N.~M. 2015, Monthly Notices
  of the Royal Astronomical Society, 451, 74

\bibitem[{{Hu} {et~al.}(2016){Hu}, {Menten}, {Wu}, {Bartkiewicz}, {Rygl},
  {Reid}, {Urquhart}, \& {Zheng}}]{Hu_2016}
{Hu}, B., {Menten}, K.~M., {Wu}, Y., {et~al.} 2016, \apj, 833, 18

\bibitem[{{Knowles} {et~al.}(1976){Knowles}, {Caswell}, \& {Goss}}]{Knowles}
{Knowles}, S.~H., {Caswell}, J.~L., \& {Goss}, W.~M. 1976, \mnras, 175, 537

\bibitem[{{Lumsden} {et~al.}(2013){Lumsden}, {Hoare}, {Urquhart}, {Oudmaijer},
  {Davies}, {Mottram}, {Cooper}, \& {Moore}}]{Lumsden_2013}
{Lumsden}, S.~L., {Hoare}, M.~G., {Urquhart}, J.~S., {et~al.} 2013, \apjs, 208,
  11

\bibitem[{{Menten}(1991)}]{1991ApJ...380L..75M}
{Menten}, K.~M. 1991, \apjl, 380, L75

\bibitem[{{Ouyang} {et~al.}(2019){Ouyang}, {Chen}, {Shen}, {Yang}, {Li},
  {Chen}, {Zhao}, \& {Sobolev}}]{Ouyang}
{Ouyang}, X.-J., {Chen}, X., {Shen}, Z.-Q., {et~al.} 2019, \apjs, 245, 12

\bibitem[{{Perley} \& {Butler}(2013)}]{2013ApJS..204...19P}
{Perley}, R.~A. \& {Butler}, B.~J. 2013, \apjs, 204, 19

\bibitem[{{Persi} {et~al.}(2011){Persi}, {Tapia}, \& {G{\'o}mez}}]{Persi}
{Persi}, P., {Tapia}, M., \& {G{\'o}mez}, M. 2011, \aap, 525, A1

\bibitem[{{Sun} {et~al.}(2014){Sun}, {Xu}, {Chen}, {Zhang}, {Wu}, {Henkel},
  {Brunthaler}, {Choi}, \& {Zheng}}]{Sun_2014}
{Sun}, Y., {Xu}, Y., {Chen}, X., {et~al.} 2014, \aap, 563, A130

\bibitem[{{Sunada} {et~al.}(2007){Sunada}, {Nakazato}, {Ikeda}, {Hongo},
  {Kitamura}, \& {Yang}}]{Sunada_2007}
{Sunada}, K., {Nakazato}, T., {Ikeda}, N., {et~al.} 2007, \pasj, 59, 1185

\bibitem[{{Szymczak} \& {Kus}(2000)}]{Szymczak_2000}
{Szymczak}, M. \& {Kus}, A.~J. 2000, \aaps, 147, 181

\bibitem[{{Szymczak} {et~al.}(2012){Szymczak}, {Wolak}, {Bartkiewicz}, \&
  {Borkowski}}]{Szymczak}
{Szymczak}, M., {Wolak}, P., {Bartkiewicz}, A., \& {Borkowski}, K.~M. 2012,
  Astronomische Nachrichten, 333, 634

\bibitem[{{Szymczak, M.} {et~al.}(2020){Szymczak, M.}, {Wolak, P.},
  {Bartkiewicz, A.}, {Aramowicz, M.}, \& {Durjasz, M.}}]{torun_exoh}
{Szymczak, M.}, {Wolak, P.}, {Bartkiewicz, A.}, {Aramowicz, M.}, \& {Durjasz,
  M.} 2020, A\&A, 642, A145

\bibitem[{{Urquhart} {et~al.}(2011){Urquhart}, {Morgan}, {Figura}, {Moore},
  {Lumsden}, {Hoare}, {Oudmaijer}, {Mottram}, {Davies}, \&
  {Dunham}}]{2011MNRAS.418.1689U}
{Urquhart}, J.~S., {Morgan}, L.~K., {Figura}, C.~C., {et~al.} 2011, \mnras,
  418, 1689

\bibitem[{{Val'tts} \& {Larionov}(2007)}]{Val'tts}
{Val'tts}, I.~E. \& {Larionov}, G.~M. 2007, Astronomy Reports, 51, 519

\bibitem[{Winkel {et~al.}(2012)Winkel, Kraus, \& Bach}]{Winkel_2012}
Winkel, B., Kraus, A., \& Bach, U. 2012, Astronomy \& Astrophysics, 540, A140

\bibitem[{{Yang} {et~al.}(2019){Yang}, {Chen}, {Shen}, {Li}, {Wang}, {Jiang},
  {Li}, {Dong}, {Wu}, \& {Qiao}}]{Yang2019}
{Yang}, K., {Chen}, X., {Shen}, Z.-Q., {et~al.} 2019, \apjs, 241, 18

\bibitem[{{Yang} {et~al.}(2017){Yang}, {Xu}, {Chen}, {Ellingsen}, {Lu}, {Ju},
  \& {Li}}]{Yang_2017}
{Yang}, W., {Xu}, Y., {Chen}, X., {et~al.} 2017, \apjs, 231, 20

\bibitem[{{Yen} {et~al.}(1969){Yen}, {Zuckerman}, {Palmer}, \&
  {Penfield}}]{Yen}
{Yen}, J.~L., {Zuckerman}, B., {Palmer}, P., \& {Penfield}, H. 1969, \apjl,
  156, L27

\end{thebibliography}

    \begin{appendix}
    \section{Additional new detections}
    While the article was being reviewed, observations were continued, and additionally seven potential new masers were noticed on the S/N around 3 or at a higher level  at 6035 MHz; details are given in the Table A.1. These seven new detections are not included in our detection statistics given at the beginning of Section 4. These sources were either not observed, or observed but not detected, by Szymczak et al. \citeyear{torun_exoh}, or as far we know by other authors. To confirm these new detections and particularly to exclude some side-lobe effects and achieve a higher S/N level, follow-up observations will be performed. A more extensive literature search for possible previous detections will be done. We aim to publish these detailed results in the next paper.
    
        \begin{table}
                \caption{List of possibly new detections. Sources observed but not detected by Szymczak et al. \citeyear{torun_exoh} are highlighted in bold. 
                Velocities in the last column are velocities of potentially detected lines.}
                \label{T4}
                \begin{tabular}{llp{0.2\linewidth}lll}
                \hline
                \noalign{\smallskip}
                 Source Name (l b) &  RA   & Dec  & V$_{lsr}$ \\
                    & (hh:mm:ss) &(dd:mm:ss)&  (km s$^{-1}$) \\
                \noalign{\smallskip}
                \hline
                \noalign{\smallskip}
                {\bf 212.06-00.74} & 06:47:13.0 & 00:26:07 & 48.2\\
                {\bf 33.645-00.227} & 18:53:32.8 & 00:31:39.0 & 60.5 \\
                {\bf 37.04-00.03} & 18:59:04.0 & 03:38:33.0 & 81.5 \\
                43.0885-00.0114 & 19:10:10.0 & 09:01:27.0 & 11.2 \\
                {\bf 45.47+00.13} & 19:14:07.0 & 11:12:16.0 & 61.0 \\
                49.405-00.370 & 19:23:30.0 & 14:26:35.0 & 55.1 \\
                49.64-0.51 & 19:24:26.0 & 14:35:34.0 & 55.3 \\
                \noalign{\smallskip}
                \hline
                \end{tabular}
        \end{table}
    \end{appendix}
\end{document}